# Benefits of Resource Strategy for Sustainable Materials Research and Development

*Christoph Helbig, Christoph Kolotzek, Andrea Thorenz, Armin Reller, Axel Tuma, Mario Schafnitzel, Stephan Krohns**

S. Krohns, M. Schafnitzel, Experimental Physics V, Center for Electronic Correlations and Magnetism, University of Augsburg, 86159 Augsburg, Germany
C. Helbig, C. Kolotzek, A. Thorenz, A. Reller, A. Tuma, Resource Lab, Institute for Materials Resource Management, University of Augsburg, 86159 Augsburg, Germany

**Material and product life cycles are based on complex value chains of technology-specific elements. Resource strategy aspects of essential and strategic raw materials have a direct impact on applications of new functionalized materials or the development of novel products. Thus, an urgent challenge of modern materials science is to obtain information about the supply risk and environmental aspects of resource utilization, especially at an early stage of basic research. Combining the fields of materials science, industrial engineering and resource strategy enables a multidisciplinary research approach to identify specific risks within the value chain, aggregated as the so-called 'resource criticality'. Here, we demonstrate a step-by-step criticality assessment in the sector of basic materials research for multifunctional hexagonal manganite $YMnO_3$, which can be a candidate for future electronic systems. Raw material restrictions can be quantitatively identified, even at such an early stage of materials research, from eleven long-term indicators including our new developed Sector Competition Index. This approach for resource strategy for modern material science integrates two objective targets: reduced supply risk and enhanced environmental sustainability of new functionalized materials, showing drawbacks but also benefits towards a sustainable materials research and development.**

## 1. Resource Strategy

The global way of life is based on intensive consumption of energy and mineral resources. Many technologies with significant socio-economic benefits require materials that are problematic due to instable, insecure or price-volatile supply [1]. Moreover, the complexity of their global supply chains leads to an increasingly precarious scenario. The sustainable extraction and use of scarce natural resources are essential tasks to reach a resource efficient techno-economic development in the future [2]. The analysis of key technologies and processes of mega sectors shows their increasing dependency on availability of strategic metals and minerals, which is often limited [3]. The whole lifecycle (e.g. extraction, processing, pre-production, production, use-phase, recycling) of raw materials goes hand in hand with significant supply risks and environmental impacts. Applying criteria, like geologic availability, geo-political dependencies, ecological compatibility and reusability of novel materials along the complete material and product lifecycle are innovative and strongly recommended directions of materials science [4,5].

More precisely, in so called mega sectors [3] like the energy sector, high technology applications, e.g., as thin-film photovoltaic for power supply [6], supercapacitors for energy storage systems or power-to-gas technology for energy transformation, implement many different elements within their functional building blocks [7], demonstrating the complexity of its upstream value chain. Scarcities or upcoming restrictions of those strategic elements [1] for essential functions like cadmium telluride utilized as p-doped semiconductor adsorber layer for light-to-energy conversion in thin-film photovoltaic systems have a strong impact on the success of those products and technologies [8]. A challenging task for modern materials science is to develop high-performance materials utilizing abundant elements to replace critical ones in existing and future technologies [5,7,8]. Therefore not only technical material parameters are essential quantities, but also the identification of raw material restrictions or benefits. Often, criticality of elements is considered first at an advanced stage of product development [8–10], during end-of-life recycling scenarios [11,12] or the concepts include only specific aspects of materials efficiency [13] or raw materials supply [14–16]. Recent comprehensive criticality studies [3,17–19] consider in detail dimensions of supply risk, environmental implications and vulnerability to supply restrictions within global, national and corporate perspectives. However, for basic materials research at an early development stage the final product made by a functionalized material is not explicitly conceivable. Only mega sectors can be addressed for a possible future application.

Here, we specify a practical guideline for materials scientists to consider criticality aspects following a multidisciplinary evaluation for the use of raw materials. Indicators within the scope of reduced supply risk and enhanced environmental sustainability were identified from literature analysis [3,17,20,21]. These indicators were evaluated by experts from the fields of material science, physics, resource strategy and economics concerning their relevance within the basic research perspective, leading to a set of eleven indicators, listed in Fig. 2 (with details in tables S1 and S2 in the Supplementary Material). All indicators of this set have a long-term and forecasting perspective, contain non-redundant information and possess adequate data quality. The newly developed Sector Competition Index (SCI) comprises the predominant raw materials consumption in mega sectors accounting for the specific value added per material input. This multidisciplinary approach serves as a guideline for materials scientists for a sustainable and more resource-efficient material development. Here, it is based on a generation of reliable data containing geographically allocated reserves, production sites and resource supply dominating countries.

We illustrate the method on a multifunctional hexagonal manganite $YMnO_3$ [22]. This compound is a promising candidate for spintronics [23], non-volatile memory materials [24], domain-wall engineered multiferroic properties [25,26] at room temperature, or the direct electrically tuned exchange bias in $YMnO_3$/permalloy heterostructures [27]. These fascinating properties open new fields for future applications due to its geometrically driven improper ferroelectric ordering [28] accompanied by a structural six fold ferroelectric domain structure exhibiting topological protected vortices [29]. Recently, high dielectric constant and appropriate loss tangents at ambient temperature have been demonstrated in these materials, allowing good prospects for $YMnO_3$ to be also used as dielectric in high power capacitors for energy storage and conversion [30].

For $YMnO_3$, we focus on the basic research stage and assume a future "virtual usability" for this compound as a functional material in electronic building blocks. Due to the negligible amount of raw material required for research activities, restrictions concerning resource availabilities rarely occur already at this stage of product lifecycle, but may become an important factor in further development stages and technology spread. Our more simplified previous approach [5] for colossal dielectric constant materials demonstrated the benefits of knowing the criticality of the raw materials at this stage to prevent or even know risks in advance. For the present approach we derive the supply risk and environmental impacts of the two elements yttrium and manganese. The development of the supply risk indicators are discussed on an annual basis from 1995 to 2013.

## 2. Materials and Product Lifecycle

For the perspective of basic materials research a holistic approach is needed [31], especially taking into account long-term and forecasting criteria for raw material supply and production [3]. Therefore a multi-level product lifecycle for an implemented material is anticipated to identify development stages and upcoming risks based on raw materials usage. These risks are expressed by manifold indicators, which comprise technological [32], geological [33], geopolitical [14], economic [34,35], social [36] and environmental aspects [37]. The progress of a technology passes specific development stages from basic research to ready-to-use product, representing the resource-based approach of the material and product lifecycle. These stages are subjected to different disciplines like material sciences, industrial engineering, resource strategy and economics.

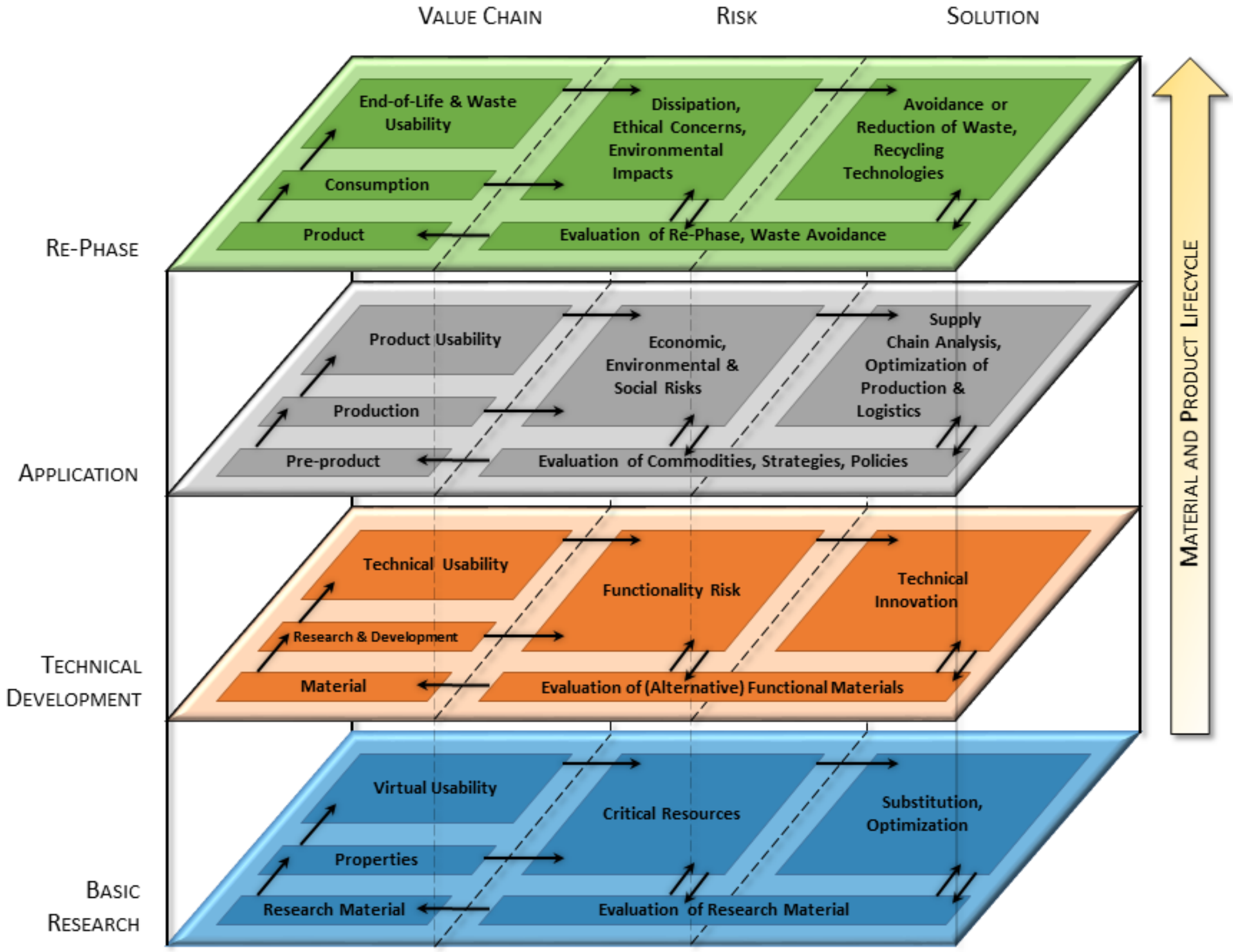

**Figure 1. Intersectional, multidisciplinary aspects of material and product lifecycles.** The four-level architecture represents basic research (blue), technical development of a material for a prototypical application (orange), application (grey) and the re-phase (green), displayed from bottom to top. Each level includes the value chain of a material or product life cycle, their concomitant risks and suitable risk mitigation strategies.

Within figure 1 we show a simplified view of various phases derived from the complex multidisciplinary and intersectional network of technology and product development: Basic research, technical development, application and re-phase. While basic research includes the conceptual functionalization of a material, in the technical development phases, the prototypical implementation for a specific product is carried out. Within the application phase the focus lies on production techniques for industrial upscaling as well as resource and energy efficiency aspects. Closing of material cycles across the whole material and product lifecycle is a necessity, thus closed-loop supply chains are established in the re-phase by recycling, remanufacturing and reuse [38,39].

The value chain in each level of figure 1 describes progress in material and product development (basic research and technology development) as well as industrial lifecycle (application and re-phase). Identification and classification of risks for all four lifecycle levels are prerequisite to develop risk mitigation strategies in order to achieve a sustainable use of functionalized resources. Many metals and metalloids show recycling rates below 1% [12]. Hence, there is potential for improvement in the design of industrial lifecycles, theoretically these materials can be recycled infinitely. Closing these material cycles would also allow for alternative material supply accompanied by reduced carbon emissions [40]. A detailed analysis of risks by combining efforts of a multidisciplinary research team, especially at the basic research level, can determine possible bottlenecks or benefits by functionalization of new materials early in a products lifecycle. A more resource-efficient use of scarce materials can be achieved or mitigation strategies developed. It is of high interest to compare criticality scores derived by this long-term approach with future criticality assessments of the same materials utilized in novel products.

Material scientists could use existing criticality assessments for a first estimate. However, all existing studies provide limited information for long-term developments. E.g., the broad coverage of metals and metalloids by Graedel and colleagues comes at the cost of only two supply risk indicators in the long-term perspective (depletion time and companion metal fraction) [18]. Other assessments either have a short- to medium-term perspective [41], a national focus [42] or only applied their method to a small set of raw materials. Therefore, we present a guideline for basic materials research on an international scale emphasizing long-term indicators.

## 3. Criticality Assessment

### 3.1 Guideline for Criticality Assessment in Basic Research

The step-by-step guideline for a resource strategy in materials science is displayed in figure 2, which represents in more detail the basic research level of figure 1. It focuses on reliable information that is accessible for material scientists. The guideline starts with an analysis of the research material requirements for the desired function and the corresponding value chain (A). The second step implements analyses of data on raw material concerning data availability and quality (B), with consideration of geographically localized information for all risk indicators. Suitable risk mitigating solutions can be assessed by calculating these indicators in the supply risk and environmental perspective (C). A detailed description for calculation of each indicator is provided in the Supplementary Material (Table S1 and S2). The guideline finishes with an interpretation and conclusion.

***Value Chain (A)***

Initially, material scientists need to become aware of material demands concerning aspects of purity of raw material or manufacturing techniques within preprocessing, in order to address inter alia specific environmental impacts or market concentration. Data analysis is either carried out on a global level or has a regional focus. For this purpose, data sources for the various indicators include scientific journal articles (like metal recycling rates [12,43]), administrative institution reports (like USGS [33]) or proprietary consultant information (like SNL Metals & Mining [44]). If necessary, data gaps can be closed by consulting a resource strategy expert.

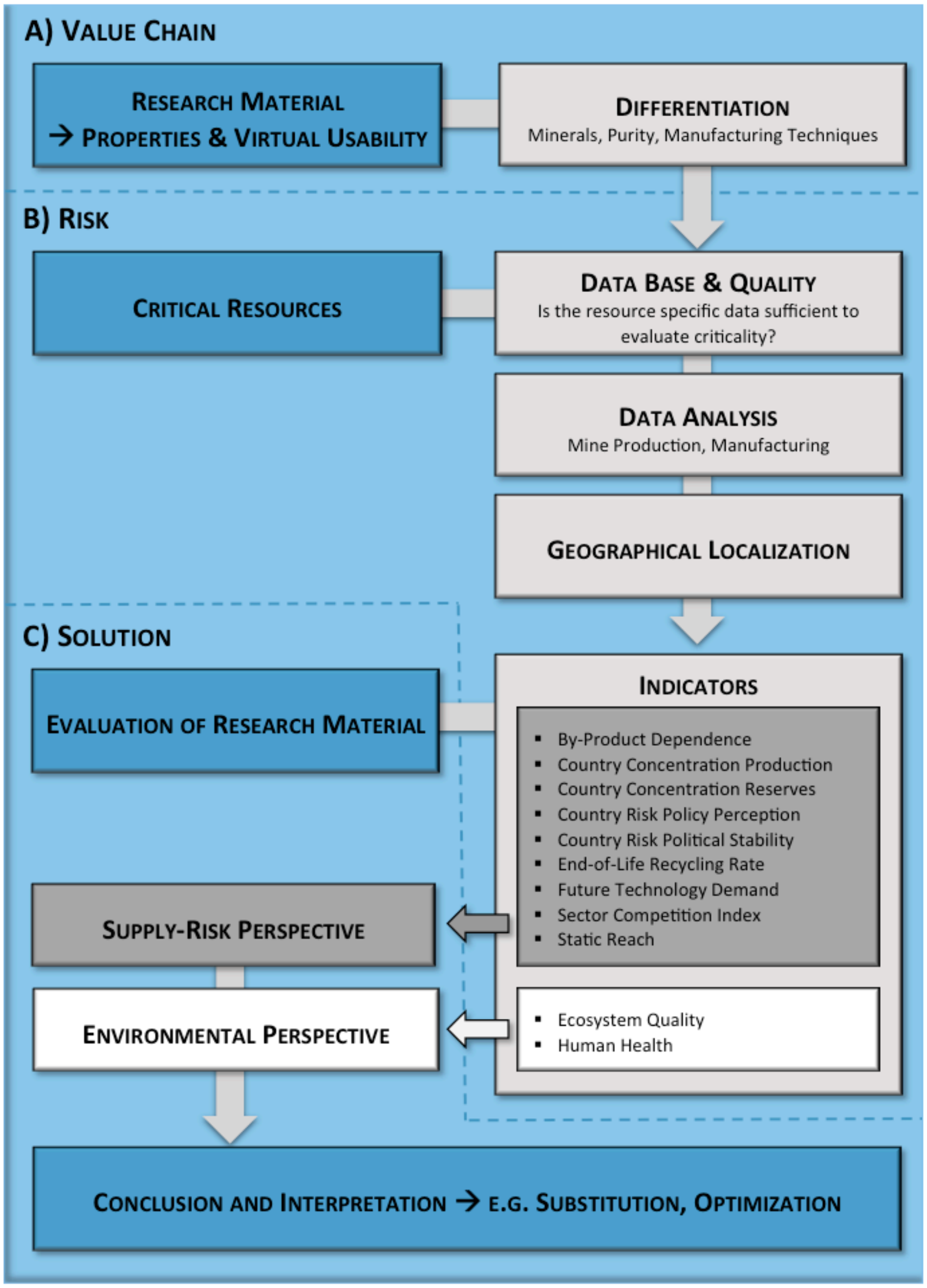

**Figure 2. Guideline for criticality assessment in materials science.** Resource-based criticality approach, starting with Value Chain (A), addressing the research material, its properties and a possible usage in a mega sector. This implies raw material and preproduction processes like purification and starting minerals. Databases are employed to determine the data set for specific Risks (B) within the value chain. Essential indicators are calculated using the data set and are grouped into two perspectives: supply risk and environmental. The aggregation of these indicators are already part of Solution (C), which allow to determine the resource criticality of used raw materials based on these quantified indicators.

### *Data Acquisition and Indicators (B)*

The eleven indicators consist of nine indicators for supply risk and two indicators for environmental impacts. Region-specific data is required for some supply risk indicators. The environmental evaluation is carried out without further geographic localization. In more detail, supply risk indicators particularly assess long-term effects concerning geological, geopolitical and technological aspects, of which most are frequently used as supply risk indicators in various criticality assessments. Geological supply risk indicators include the static reach of reserves and by-product dependence. The static reach of reserves calculates the ratio between reserves estimations and annual primary production [45]. The by-product dependence is calculated as the share of primary production that originates from mines, which have other host minerals [16]. Geopolitical aspects are covered by the country concentration including both reserves and production as well as country risks concerning policy perception [46] and political stability [14]. Country and company concentrations are calculated as the Herfindahl-Hirschman-Index (HHI) of the mining or refining activities of producing countries or companies. The country risk policy perception is determined as an average of the policy perception of those producing countries, weighted by the primary production shares. Considering a country-average of the Political Stability and Absence of Violence/Terrorism meta-indicator from the Worldwide Governance Indicators, again weighted by the primary production shares [14] leads to the country risk political stability. Technological supply risk aspects take into account future technology demand [32] and end-of-life recycling rate [12]. The demand increase due to future technologies is the ratio between estimated additional 2030 demand from identified future technologies in comparison to 2006 production volumes [32]. End-of-life recycling rate is the ratio between material recycled from old-scrap and discarded material. Additionally, the new Sector Competition Index measures the average value added per mass flow weighted by mega sector application shares, which is further described in section 3.2. The environmental perspective follows the ReCiPe method (v1.08) in its endpoint categories Human Health and Ecosystem Quality [47] taken from the ecoinvent (v2.2) life cycle database [48].

***Criticality Assessment – Solution (C)***
The interpretation of criticality indicators includes the installation of thresholds for each indicator and the harmonization of data scales. Each indicator is normalized to a score between 0 and 100, as described in the Supplementary Material (Table S1 and S2). For the supply risk dimension a total score is achieved by equal weighting of each indicator. Results of two alternative weighting options are displayed in the Supplementary Material (Table S3 and Figure S2). 'Hierarchist' normalization and 'average' weighting for the European region are used for all environmental impact data [47]. Mitigation strategies, like improved resource efficiency or material substitution, are derived and depend on criticality assessment, especially focusing on individual indicator results (see also Tables S1 and S2).

## 3.2 Sector Competition Index

The Sector Competition Index is a new developed indicator of key relevance for the supply risk indicator set addressing the problem of competing raw material demand from specific industrial sectors. From a material scientist's perspective, it is important to know if the required raw material is predominantly used in sectors that are able to pay high resource prices. This is expressed by a higher ratio between value added and raw material input per sector, defining the sector resource productivity. Competing against highly productive mega sectors in terms of demand implies a supply risk, as these sectors are able to secure prioritized supply of raw materials because of their high specific value added. Figure 3 displays the value added per material input of 17 mega sectors identified in the EU study on critical raw materials for the EU of 2014 [3]. The sector resource productivity $P_s$ of a sector s is the value added $VA_s$ per total mass of utilized material $m_s$, which is the sum of all resources $m_{rs}$ in this sector:

$$P_s = \frac{VA_s}{m_s} = \frac{VA_s}{\sum_r m_{rs}} \qquad \text{(Eq. 1)}$$

The least productive sector according to the EU study [3] data has been construction material, closely followed by metal applications. The highest productivities have been identified for the applications refining and pharmaceuticals as well as electronics and information and communication technology (ICT).

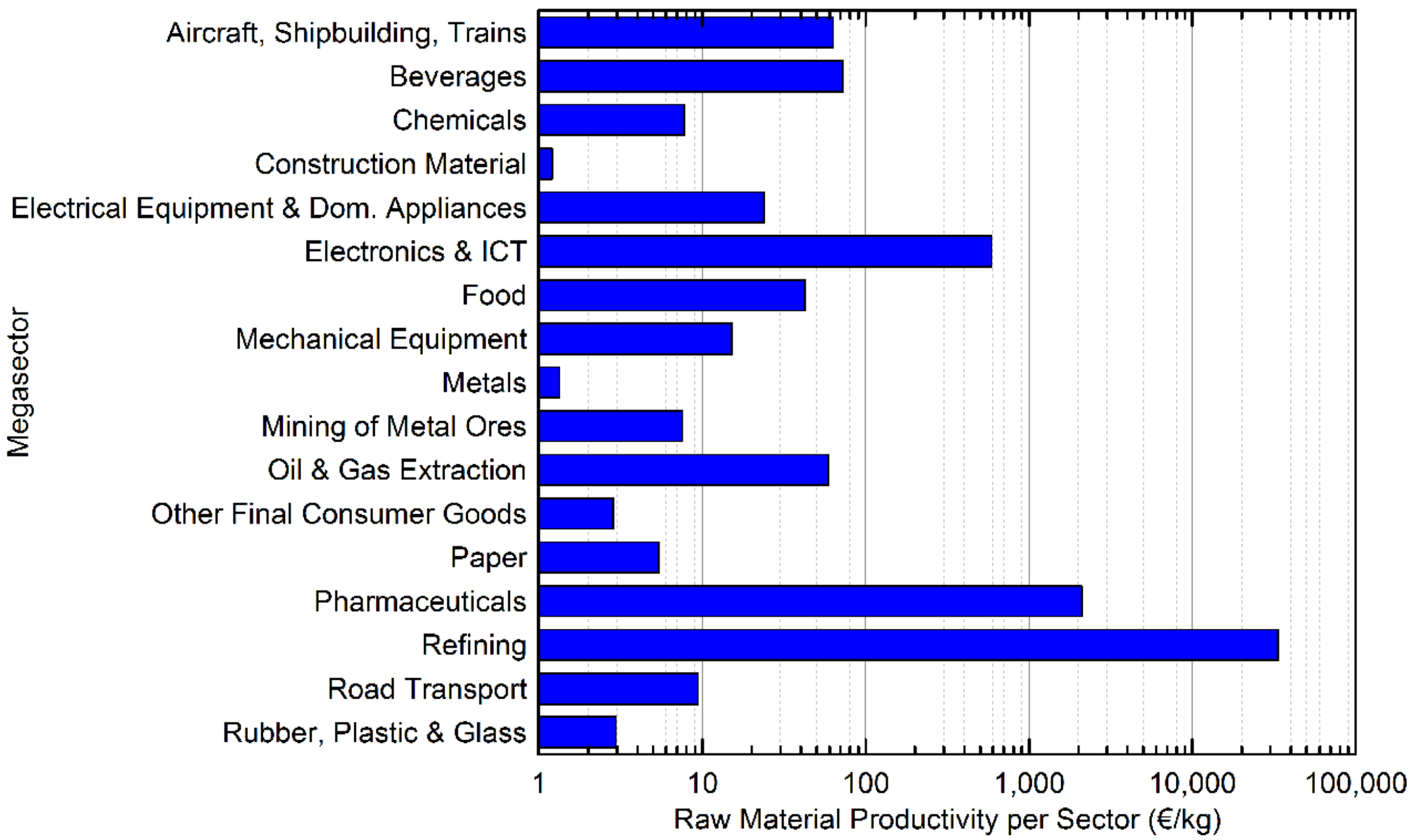

**Figure 3. Sector-specific resource productivities for year 2010.** Mega sectors displayed versus the logarithmic raw material productivity (€/kg). Productivity has been calculated as value added of each megasector devided by total material input, according to the European Commission [3].

In order to calculate a resource-specific indicator, the Sector Competition Index ($SCI_r$) is calculated from rescaled resource productivities of the sectors, weighted by application share (the ratio between mass of a resource used in a sector $m_{rs}$ and total production of the resource $m_r$). The rescaling of a normalized productivity $P_s^*$ assumes a logarithmical relation between criticality of a resource, expressed on a scale from 0 to 100 and resource productivity $P_s$ of its application sectors. Therefore the least productive sector is set to 0 and the most productive sector to 100:

$$SCI_r = \sum_s \frac{m_{rs}}{m_r} P_s^* = 100 \sum_s \frac{m_{rs}}{m_r} \frac{\log \frac{P_s}{\min_s P_s}}{\log \frac{\max_s P_s}{\min_s P_s}} \qquad \text{(Eq. 2)}$$

The Sector Competition Index is calculated for 54 major materials (more details are in Figure S1). The highest scores are derived for indium, gallium, germanium and tellurium, exceeding values of 40, all of them frequently used in the electronics and ICT industry. The lowest scores, all below 2, are for chromium, coking coal, gypsum, nickel, sawn softwood, which are mass metals or non-metal raw materials.

## 4. Results and Discussion of Resource Criticality of $YMnO_3$

The resource-strategic criticality assessment considers potential obstacles, side-effects and impacts of material utilization and accompanies the product and material development. Following the guideline presented in figure 2 and section 3.1, we apply it to both embodied metals of hexagonal manganite: rare earth element yttrium and transition metal manganese. For both elements necessary databases are analyzed, gaining sufficient resource-, and region-specific information to quantitatively evaluate the supply risk and environmental perspective.

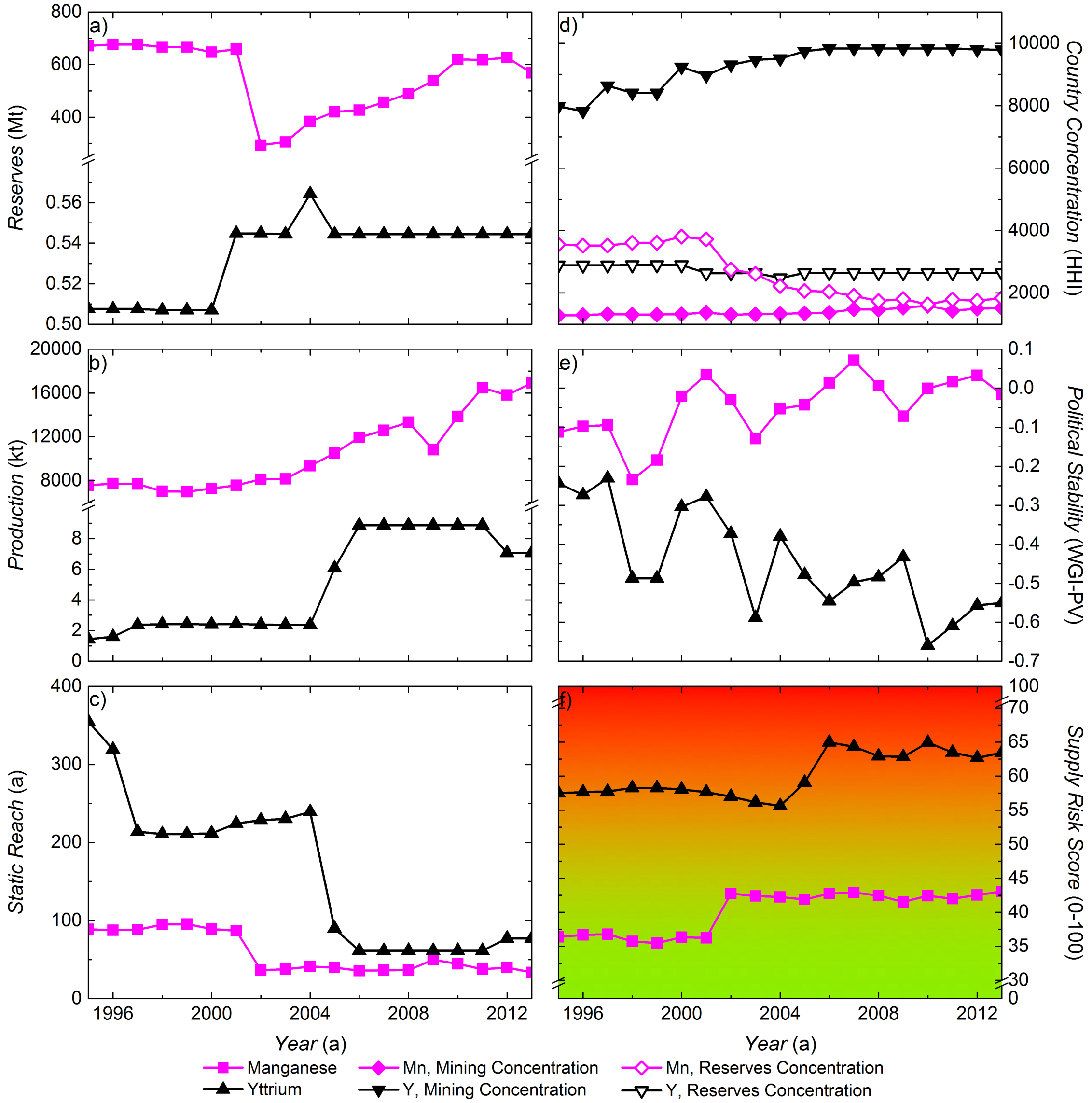

**Figure 4. Time dependent development of criticality scores of Yttrium and Manganese from 1995 to 2013.** a) – e) show time dependent raw materials data of Reserves, Production, Static Reach, Country Concentration and Political Stability. The development of the total supply risk scores (equal weighting) of Yttrium and Manganese from 1995 to 2013 is shown in f), the colour code visualizes the criticality score of supply risk ranging from green (not critical) to red (critical).

Trends and consistency of the derived criticality indicators are displayed by their historic development. Figure 4 shows the timeline for both considered elements for various indicators: reserves (a), mining production (b), resulting static reach of reserves (c), country concentration mining and reserves (d), political stability of mining countries (e) and the total supply risk scores (equal weighting) for the time span from 1995 to 2013 (f). The data in a) – e) are in units as collected from data sources, prior to criticality determination, allowing for full interpretation. The environmental implications are not reevaluated on an annual basis within the Ecoinvent database. In particular, for yttrium and manganese, only a single data set for 2010 is available.

The development of the supply risk scores exhibits an almost constant time-dependent behavior except for two step-like increases from 2004 to 2006 for yttrium and from 2001 to 2002 for manganese, respectively (figure 4f). The increase for yttrium is explained by quadrupled primary production, which is depicted in figure 4b. This production was mainly covered by China, leading to both a decreased static reach of Yttrium (4c) and a slight further increase of the country concentration (4d). For manganese, the increase in criticality originates from a reevaluation of South African reserves in 2002 [49] (c.f. figure 4a), resulting in a significantly decreased static reach (4c), which changed from 87 years in 2001 to 36 years in 2002. Further deviations from a constant behavior in reserves and production (4a, 4b) compensate each other with no further impact on the static reach. E.g. the manganese primary production volumes and reserves outside of South Africa have simultaneously gradually increased. In addition, the mentioned reevaluation of South African reserves lowered the country concentration of reserves (4d). The time-dependent change of political stability of production countries for both elements influences the supply risks only marginally.

All eleven indicators for manganese and yttrium are listed in figure 5: nine for supply risk for 2013 and two for environmental perspective for 2010, respectively. Equal weighting is used in each perspective to identify the total criticality scores of both yttrium and manganese. In the Supplementary Material (Table S3 and Figure S2) we show a sensitivity analysis considering two alternative weighting options: (i) equal weighting of four indicator groups and (ii) higher weighting of single indicators. The total criticality scores of each material differ due to the weighting options by a maximum of five points. For yttrium, the highest criticality scores are derived due to the nearly monopoly situation of the refining production of China and its corresponding political evaluation (Country Concentration Production). This leads to high market concentration accompanied by rather poor values for Policy Perception and Political Stability. Yttrium lacks in end-of-life recycling for 2013, giving a maximum criticality in this indicator. Ciacci and colleagues assume a future potential end-of-life recycling rate of 18%, especially from phosphor powders and garnet lasers [50]. The supply risk of yttrium is lowered by its estimated Future Technology Demand and rather low Sector Competition Index due to main use in mega sectors of little value added, e.g. chemicals industry (c.f. figure 3), Furthermore, high yttrium reserves lead to a long Static Reach. Even though rare earth mining is frequently accompanied by uranium and thorium ores, the overall specific environmental implications are interestingly evaluated as non-critical [51], despite the ongoing discussions on toxicology and environmental regulations [52]. Therefore, from a material scientist's perspective, future yttrium applications should be designed for recycling and a more widespread production should be pursued.

Compared to yttrium, manganese has a lower supply risk score, which mainly emerges from a reduced By-Product Dependence (manganese is generally a host metal), an even lower estimated Future Technology Demand and main uses in mega sectors with low value added, particularly in the metals sector [2] (c.f. figure 3). Higher supply risks are based on country concentration (mainly South Africa) and a rather short static reach (34 years). The specific environmental implications for manganese are of minor importance due to relatively high abundance of manganese in the earth's crust [53] and therefore higher ore grades and low-hazard extraction technologies [48]. Therefore, future manganese applications should particularly strive for high resource-efficiency in order to reduce overall material demand. Incentives for further development of extraction and refining technologies are required to access techno-economically more resources enhancing the static reach.

| Perspective | Indicator | Dimension | Ytrrium | | | Manganese | | |
|---|---|---|---|---|---|---|---|---|
| | | | Value | Score (0 - 100) | | Value | Score (0 - 100) | |
| Supply Risk | Companion Metal Fraction | % | 84.30 | 84.30 | 63.43 | 0.00 | 0.00 | 43.03 |
| | Country Concentration Production | HHI | 9795.32 | 99.64 | | 1521.06 | 67.05 | |
| | Country Concentration Reserves | HHI | 2646.52 | 76.74 | | 1840.78 | 70.38 | |
| | Country Risk Policy Potential | dimless | 21.49 | 78.51 | | 46.36 | 53.64 | |
| | Country Risk Political Stability | WGI-PV | -0.55 | 61.01 | | -0.02 | 50.33 | |
| | Demand Incr. due to Fut. Tech. | dimless | 0.01 | 1.00 | | 0.00 | 0.00 | |
| | Recycling Rate | % | 0.00 | 100.00 | | 53.00 | 47.00 | |
| | Sector Competition Index | dimless | 32.50 | 32.50 | | 14.60 | 14.60 | |
| | Static Reach Reserves | year | 76.97 | 37.20 | | 33.59 | 84.25 | |
| Environmental | Human Health | endpoints (per kg) | 0.69 | 6.60 | | 0.20 | 1.98 | |
| | Ecosystem Quality | | 0.45 | | | 0.06 | | |

**Figure 5. Criticality scores of Yttrium and Manganese.** Values for the 11 indicators of the criticality assessment. Data for supply risk is for 2013, environmental data for 2010. The color code visualizes the impact of the criticality score on materials usage for each perspective, ranging from green (not critical) to red (critical).

## 5. Conclusion

In this article, we describe a guideline for material scientists to implement resource strategy considerations in basic research of the development of functional materials. Material and product lifecycle consists of four stages: basic research, technical development, application and re-phase. Our guideline empowers material scientists to evaluate supply risk and environmental aspects based on raw materials. Eleven quantitative indicators utilized for this purpose allow for a holistic evaluation, embedded into the comprehensive resource strategy, that enables a long-term and future-oriented assessment. The newly developed Sector Competition Index demonstrates its importance for value added of raw material input used in mega sectors. We use this approach to analyze the resource-based prospects of a promising multifunctional material, $YMnO_3$. The supply risk is evaluated for a time span of 20 years, demonstrating that for both considered elements, criticality in general remains stable over time. Possible exceptions are reevaluation of reserves or external effects like significant changes in production and consumption. Therefore, the present criticality assessment is applicable for the long-term perspective of basic research. In particular, the equally weighted indicators give rise to a moderate to high supply risk of yttrium and manganese. Conversely, environmental implications of yttrium and manganese are of low criticality. Benefits are created by identifying possible risk mitigation strategies. The time-to-market process for $YMnO_3$ functionalized in future products can be accelerated if for example diversified supply and design-for-recycle are considered already at basic research level, especially for yttrium. Environmental regulations for either element seem unlikely.

Implementing resource strategy concepts already at basic research stage will help to identify promising sustainable materials, improving the resource efficiency to an enhanced sustainable circular economy. This allows developing technologies that reduce the dissipative usage of scarce natural resources. It's a challenge to retrospectively derive the time-dependent evolution of materials' criticalities for all four stages of the material and product lifecycles to prove the impact of single indicators and benefits of risk mitigation strategies.

# Benefits of Resource Strategy for Sustainable Materials Research and Development

## Supplementary Information

**C. Helbig[1], C. Kolotzek[1], A. Thorenz[1], A. Reller[1], A. Tuma[1], M. Schafnitzel[2], S. Krohns[2,*]**

[1]Resource Lab, University of Augsburg, 86159 Augsburg, Germany
[2]Experimental Physics V, Center for Electronic Correlations and Magnetism, University of Augsburg, 86159 Augsburg, Germany

*e-mail: stephan.krohns@physik.uni-augsburg.de

### INDICATORS FOR CRITICALITY ASSESSMENT

**Table S1. Indicators for the Economic Perspective of the Criticality Assessment.**

| **Indicator** | **Description** | **Calculation (Normalization: 0 to 100 scale)** | **Data Source** | **Examples of Positive Impacts on Indicators** |
|---|---|---|---|---|
| By-Product Dependence | The by-product dependence is the percentage of the element mined as a by-product of the global production of another element. This is the case, when mining solely for the raw material itself is not economically feasible.<br><br>This figure is a measure of the potential inability to increase primary production in response to an increase in demand. | Share of host metal mine production in total primary production in % | [1,2] | - Development of minor metal mining projects<br>- New extraction technologies for minor metals |
| Country Concentration Production (CCP) | The concentration of annual production of a raw material at the country level is measured by the Herfindahl-Hirschman Index (HHI), which is the sum over the squares of the production shares of the countries in percent. HHI values range from 0 to 10000, the normalization numbers 17.5 and 61.18 are set to fix an HHI value of 1800 to the supply risk score of 70 and the maximum criticality to 100 [3].<br><br>The value indicates directly market concentration in a few countries and thus the possibility of strategic exploitation of a monopolistic position at times of international crisis or dispute. | $CCP = 17.5 \times \ln(\text{HHI}) - 61.18$ | [4] | - Development of mining projects in minor producing countries |

| **Indicator** | **Description** | **Calculation (Normalization: 0 to 100 scale)** | **Data Source** | **Examples of Positive Impacts on Indicators** |
|---|---|---|---|---|
| Country Concentration Reserves (CCR) | The concentration of estimated reserves of a raw material at the country level is measured by the Herfindahl-Hirschman Index (HHI), which is the sum over the squares of the reserve shares of the countries in percent. HHI values range from 0 to 10000, the normalization numbers 17.5 and 61.18 are set to fix an HHI value of 1800 to the supply risk score of 70 and the maximum criticality to 100 [3].<br><br>The value indicates possible future market concentration in a few countries and thus the possibility of strategic exploitation of a monopolistic position at times of international crisis or dispute. | $\boldsymbol{CCR} = \mathbf{17.5} \times \mathbf{ln}(\text{HHI}) - \mathbf{61.18}$ | [4] | - Ore exploration<br>- New extraction technologies |
| Country Risk Policy Perception (CRPP) | The indicator Policy Perception (PPI) is an assessment of the ability of producing countries to implement new mining projects, weighted by the production share (P) in each country. The Policy Perception is evaluated by mining industry experts and summarized by the Fraser Institute.<br><br>The value is a measure of the ability of the market further increase production based on the rule of law and governance procedures in producing countries. | $\boldsymbol{CRPP} = \sum_{c} (\mathbf{100} - \text{PPI}_{c}) \times \boldsymbol{P_c}$ | [5] | - Development of mining projects in countries with good policy perception<br>- Increase of policy perception in main mining countries |
| Country Risk Political Stability (CRPS) | The risk of political instability in producing countries is measured by the Worldwide Governance Indicator for Political Stability and Absence of Violence/Terrorism (WGI-PV), presented by the World Bank, weighted by the production share in each producing country (P). The values are normalized linearly taking into account that WGI scores range from -2.5 to 2.5.<br><br>The value is an indication of the likelihood of disruption in production and export in the countries concerned due to unrest, coups d'état, terrorism or other situations involving violence. | $\boldsymbol{CRPS} = \mathbf{20} \sum_{c} (\mathbf{2.5} - \text{WGI-PV}_{c}) \times \boldsymbol{P_c}$ | [6] | - Development of production projects in politically stable countries<br>- Stabilization of main producing countries |
| Recycling Rate (End-of-Life) | The end-of-life recycling rate (EoL-RR) of a raw material is measured by the ratio of current annual recycled material flow to the annual discard rate of the raw material.<br><br>The value gives an estimate of the amount of available secondary material, which is independent of mining and primary refining activities and can thus smooth out supply disruptions or price peaks. | 100 - EoL-RR (in %) | [7] | - Development of new recycling technologies<br>- Increase of scrap collection rate |
| Future Technology Demand | Future technology demand is given by the ratio of expected additional demand in a future year due to new, future technologies and global production in a past year.<br><br>The value gives an indication of the market pressure for increasing global extraction due to future technologies and therefore of potential additional competition in the commodity markets. | Future Technology Demand in % (capped at 100) | [8] | - Diversification of high-tech metal demand<br>- Material efficiency for future technologies |

| Indicator | Description | Calculation (Normalization: 0 to 100 scale) | Data Source | Examples of Positive Impacts on Indicators |
|---|---|---|---|---|
| Sector Competition Index | Resource productivity of the application sectors of a raw material, in order to estimate their competitiveness concerning higher commodity prices. | see Sector Competition Index of main text | [9] | - Increase resource productivity of megasectors with low raw material productivity |
| Static Reach Reserves | The static reach (SR) of the reserves of a raw material is measured by the ratio of annual primary production to the estimated global reserves. Reserves are ores currently technically and economically extractable from known deposits. Normalization numbers 1/5 and 8/1000 are set to give a static reach of 100 years the lowest score of 0 and a static reach of 50 years the score of 70 [3].<br><br>The value gives an indication of the market pressure for further exploration and for the development of new extraction capabilities, possibly leading to higher price levels. | $\boldsymbol{SR_{reserves}} = \mathbf{100} - \frac{\mathbf{1}}{\mathbf{5}}\times\mathrm{SR} - \frac{\mathbf{8}}{\mathbf{1000}}\times\mathrm{SR}^{\mathbf{2}}$ | [4] | - Ore exploration<br>- Development of new extraction technologies |

**Table S2. Indicators for the Environmental Perspective of the Criticality Assessment**

| Indicator | Description | Calculation | Data Source | Examples for Positive Impacts on Indicators |
|---|---|---|---|---|
| Ecosystem Quality | Life-Cycle Impact on Ecosystem Quality (EQ) through biodiversity loss or extinction of species according to the cradle-to-gate Life Cycle Inventory. Normalization is set to transform ReCiPe endpoints of metals from the Ecoinvent database to values between 0 and 100 [3].<br><br>Impacts include Agricultural Land Occupation, Climate Change, Freshwater Ecotoxicity, Freshwater Eutrophication, Marine Ecotoxicity, Natural Land Transformation, Terrestrial Acidification, Terrestrial Ecotoxicity and Urban Land Occupation | ReCiPe v2.2 Endpoint Ecosystem Quality,<br>Ecol $= \mathbf{log_{10}}(\mathrm{HH} + \mathrm{EQ} + \mathbf{1})\times\mathbf{20}$ | [2,10] | - Reduction of energy requirements and emissions during mining activities<br>- Compensatory measures |
| Human Health | Life-Cycle Impact on Human Health (HH) through injuries or death according to the cradle-to-gate Life Cycle Inventory. Normalization is set to transform ReCiPe endpoints of metals from the Ecoinvent database to values between 0 and 100 [3].<br><br>Impacts include Climate Change, Human Toxicity, Ionising Radiation, Ozone Depletion, Particulate Matter Formation and Photochemical Oxidant Formation | ReCiPe v2.2 Endpoint Human Health,<br>Ecol $= \mathbf{log_{10}}(\mathrm{HH} + \mathrm{EQ} + \mathbf{1})\times\mathbf{20}$ | [2,10] | - Reduction of energy requirements and emissions during mining activities<br>- Compensatory measures |

## SECTOR COMPETITION INDEX

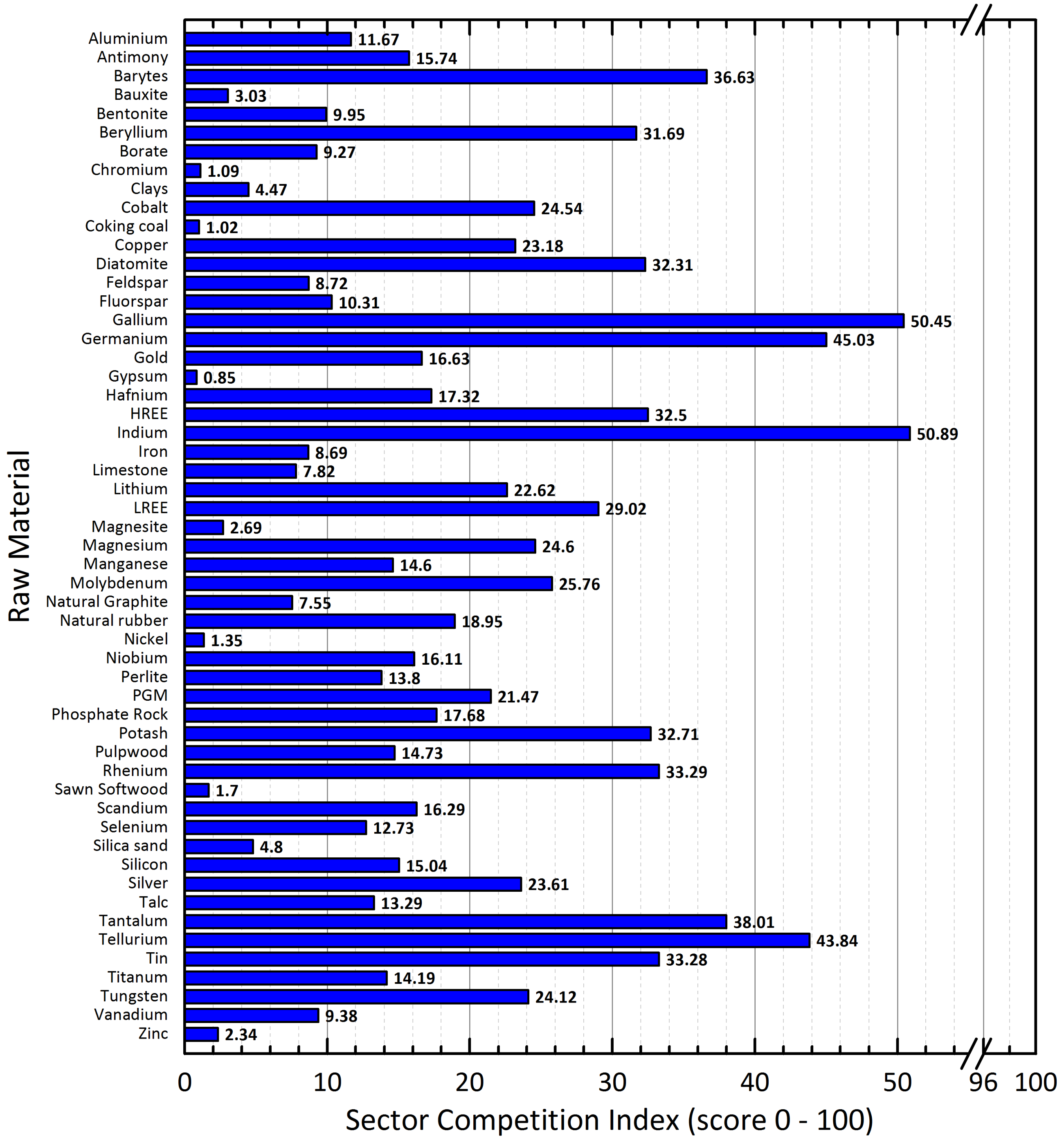

**Figure S1. Results for the Sector Competition Index for 54 major materials in 2010.** Sector competition index implies for each raw material the value added per raw material input for 17 megasectors in the EU (cf. figure 3 of the main text). The criticality indicator is calculated considering the raw material productivity per sector (€/kg) times the used annual production fraction within these megasectors. High criticiality scores indicate a high raw materials productivity, i.e. spice metals, e.g., rare-earth-elements for high-tech application. Low values implies mass material usage or application in less economically dominating megasectors.

## SENSITIVITY ANALYSIS

**Table S3. Weightings for the sensitivity analysis of the supply risk assessment.** Equal weighting: Each of the nine indicators gets the same weight. Group weighting: Indicators are grouped into equally weighted categories in analogy to Helbig et al. (2016) [11] with categories (a) supply reduction risk, (b) demand increase risk, (c) concentration risk, (d) political risk. Single weighting: Indicators used by Graedel et al. (2012) [3] for a long-term perspective and the newly developed Sector Competition Index are given double weighting.

| Supply Risk Indicator | Yttrium | Manganese | Equal Weighting | Group Weighting | Single Weighting |
|---|---|---|---|---|---|
| By-Product Dependence | 84.30 | 0.00 | 11.11% | 8.33% (b) | 16.67% |
| Country Concentration Production | 99.64 | 67.05 | 11.11% | 12.50% (c) | 8.33% |
| County Concentration Reserves | 76.74 | 70.38 | 11.11% | 12.50% (c) | 8.33% |
| Country Risk Policy Perception | 78.51 | 53.64 | 11.11% | 12.50% (d) | 8.33% |
| Country Risk Political Stability | 61.01 | 50.33 | 11.11% | 12.50% (d) | 8.33% |
| End-of-Life Recycling Rate | 100.00 | 47.00 | 11.11% | 12.50% (a) | 8.33% |
| Future Technology Demand | 1.00 | 0.00 | 11.11% | 8.33% (b) | 8.33% |
| Sector Competition Index | 32.50 | 14.60 | 11.11% | 8.33% (b) | 16.67% |
| Static Reach Reserves | 37.20 | 84.25 | 11.11% | 12.50% (a) | 16.67% |
| Total Supply Risk Yttrium | | | 63.43 | 66.45 | 61.69 |
| Total Supply Risk Manganese | | | 43.03 | 47.80 | 46.10 |

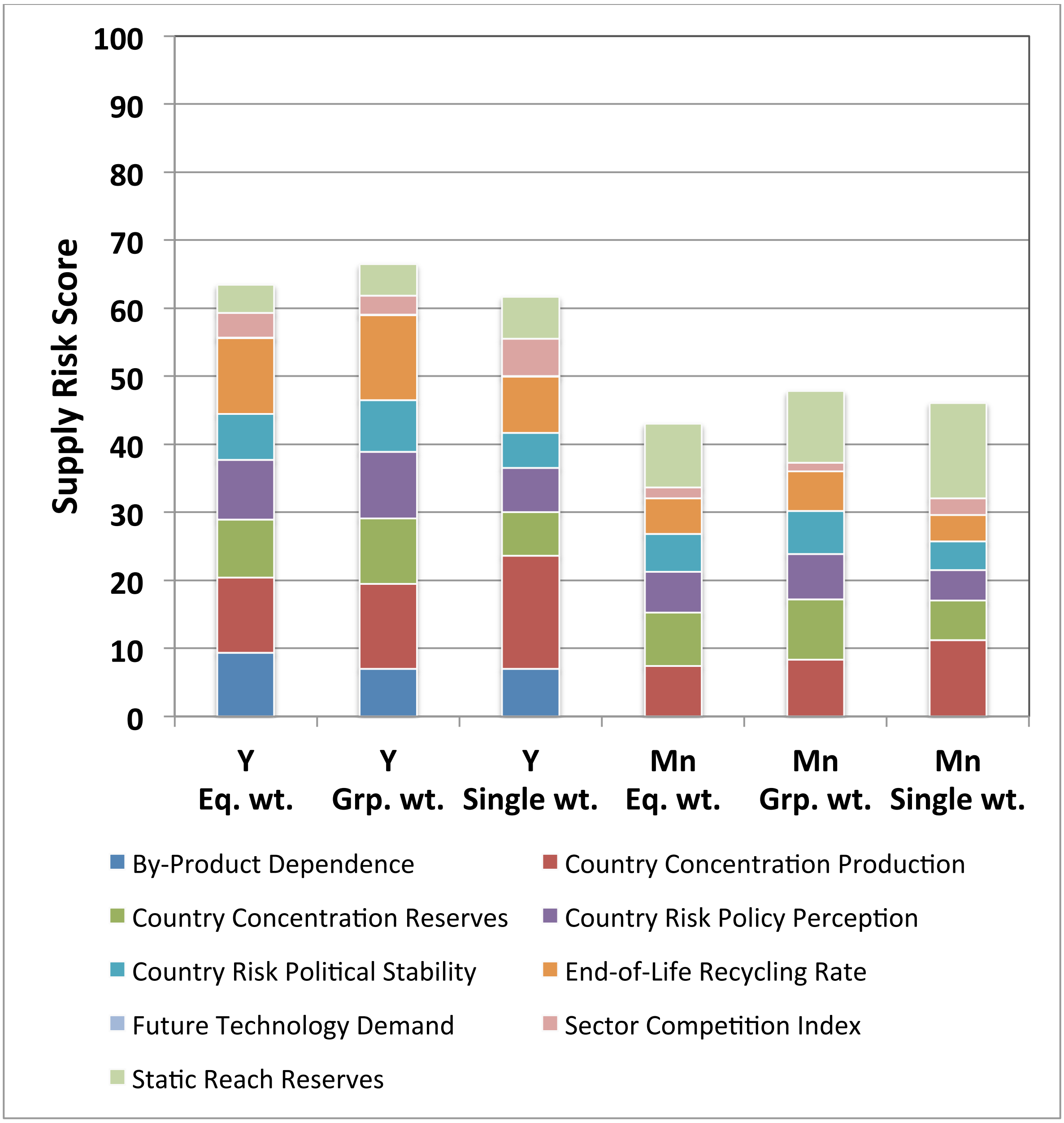

**Figure S2. Results of the sensitivity analysis for supply risk with alternative weightings.** Eq. wt.: Equal weighting as used in the main article. Grp. wt.: Group weighting concerning the categories supply reduction risk, demand increase risk, concentration risk and political risk. Single wt.: Single weighting with doubled importance of by-product dependence, sector competition index and static reach reserves. Specific indicator weightings are displayed in table S3.